\documentstyle[prl,preprint,aps,epsf]{revtex}

\begin{document}
\draft
\title{Effective Lagrangians and Parity-Conserving \\ Time-Reversal
Violation at Low Energies}
\author{Jonathan Engel and Paul H. Frampton}
\address{Department of Physics and Astronomy,
University of North Carolina, \\ Chapel Hill, North Carolina
27599-3255}
\author{Roxanne P. Springer}
\address{Department of Physics, Duke University,
Durham, North Carolina  27708}
\date{\today}
\maketitle

\begin{abstract}

Using effective Lagrangians, we argue that any time-reversal-violating but
parity-conserving effects are too small to be observed in flavor-conserving
nuclear processes without dramatic improvement in experimental
accuracy.  In the process we discuss other arguments that have
appeared in the literature.

\end{abstract} 
\pacs{PACS-1995:  11.30.Er, 11.10.E, 24.80.D}

\narrowtext
 
The discrete spacetime symmetries of parity (P) and time reversal (T) have
played a crucial role in our understanding of fundamental interactions.
Parity violation is a general feature of weak interactions,
observed in a wide range of phenomena.  By contrast, time-reversal violation 
has been seen only in the neutral kaon system.  Yet
measurements in the kaon system alone are insufficient to determine
whether the Kobayashi-Maskawa (KM) \cite{KM} mechanism of explicit T
violation is operating or whether extra-standard-model physics is at play.
Even the B-meson factories under construction may not be able to tell us if
the source of T-violation is really the KM mechanism.

The possibility that T-violation might arise outside the standard model has
motivated a number of recent low-energy (MeV-range or less) experiments.
These measurements, which do not test KM-based T violation but may be
sensitive to other sources, are classified according to whether or not the
measured observables violate P as well as T\cite{gould}.  Electric-dipole
moments, both of elementary particles and atoms, are T-violating and
P-violating (TVPV) observables.  The quantities we focus on here are
T-violating but P-conserving (TVPC) and flavor-conserving.  They include
correlations both in $\gamma$-decay\cite{gamma} and neutron
scattering\cite{Koster} as well as quantities extracted from nuclear tests
of detailed balance\cite{db}.  Some observables in beta decay\cite{beta} are
TVPC but are flavor changing, and will not be considered
here.

The reason the TVPC experiments\cite{gould} are interesting is that limits on
the quantities they measure are still quite weak (much weaker than the limits
on similar TVPV quantities), raising the possibility that TVPC effects could
be relatively large.  The experiments are not very sensitive in part because
of the inability of a single pion, which is largely responsible for the 
strong
force between nucleons in a nucleus, to transmit a TVPC
interaction\cite{Simonius}.  Though the experiments are improving, the best
published limit on the effective TVPC coupling of the nucleon to the $\rho$,
the lightest relevant meson, is still only about $10^{-2}$ times the normal
strong $\rho N N$ coupling\cite{kind1,kind2}

Is it possible that large T violation from outside the standard model is
lurking just below current limits, that a TVPC effect could appear at
$10^{-3}$ or $10^{-4}$ times the strong coupling $g$?  
Prior work has addressed this and related issues. 
Herczeg et 
al.\ \onlinecite{Herczeg} have shown that in any 
renormalizable gauge theory, with the $\theta_{QCD}$-term neglected, 
Feynman graphs representing a TVPC flavor-conserving
quark-quark interaction must contain more than two non-QCD non-QED
vertices. This theorem suggests that low-energy TVPC effects
will be strongly suppressed in theories that generate T-violation through the
weak coupling of heavy bosons to quarks.  In other kinds of models, however,
the theorem leaves the issue open.  If, for example, the
bosons that break T are strongly coupled and confined, the TVPC interaction 
may not be perturbative, and an examination of its graphical structure may 
not 
provide constraints.  
Conti and Khriplovich\cite{CK} have approached the issue in a different way, 
arguing that measured limits on electric dipole moments, which are TVPV, 
constrain TVPC vertices through graphs that also contain parity-changing Z 
bosons.  They obtain a limit of $10^{-10}$ for the ratio of TVPC to
strong interactions.  Here we argue that TVPC effects are small but not
necessarily that small.  More specifically, under very conservative 
assumptions,
regardless of how time-reversal invariance is broken, effective TVPC 
couplings
lie at or below $10^{-8}$ times the strong coupling $g$.

To address these matters in a systematic fashion, we draw on effective field 
theory, which makes possible very general conclusions about low-energy 
phenomena despite our
ignorance of physics at high energies\cite{Leff}.  Explicit dependence
on any physics at scales
much larger than a typical momentum transfer in a low-energy experiment
can be removed from the full theory.  This
leaves an effective Lagrangian consisting of a 
sum of nonrenormalizable 
(dimension greater than four) operators 
involving standard-model fields.  The effects of the high-energy
physics
appear in factors multiplying each
of the operators in the effective Lagrangian.  If the unknown physics
is associated with some high-energy scale
$\Lambda$, the effective Lagrangian that represents its low-energy limit 
can be written as
\begin{equation} 
\label{e:L} 
{\cal L}_{eff} = {\cal L}_0 + \frac{1}{\Lambda}
{\cal L}_1 + \frac{1}{\Lambda^2} {\cal L}_2 + \ldots \, \, ,
\end{equation}
where each ${\cal L}_i$ contains a series of operators of dimension $i+4$,
each multiplied by a dimensionless coefficient expected to be of order one. 
[Enhancements of coefficients beyond the expectations of ``naive dimensional 
analysis"\cite{har} are possible however; an example is
the $\Delta I =1/2$ rule, where the coefficient responsible
is about 20.] 
The results of any calculation using this Lagrangian will be a power
series in $(p/\Lambda)$, where $p$ is a typical momentum transfer in the
process under consideration.  If $p << \Lambda$, the series will in general
converge, and truncating it at any order in $1/\Lambda$ provides a  
theory with a finite number of terms, predictive to that order in 
$1/\Lambda$.

To use this formalism to bound low-energy TVPC effects without knowing
anything about physics at higher scales, we must first identify a maximum 
quark/gluon momentum $p$ relevant to the experiments we consider and a 
minimum scale
$\Lambda$ with  which T-violating physics could be associated.  To be
conservative we choose the proton mass, $m_p \approx $ 1 GeV, 
to be the
momentum scale $p$.  The appropriate value for $\Lambda$ is less obvious, but
it should be at least the mass of the $Z^0$.  Undiscovered gauge particles
could conceivably be lighter, but precision tests of the standard model
impose strong constraints on the couplings and/or masses such particles can
have.  The contribution of any gauge boson of mass $m$ coupled to up and down
quarks ($q$) with strength $f$ will contain a factor of $f^2/m^2$, which must
be smaller than the analogous factor for $Z^0$ exchange to avoid conflict
with the measurement of the partial width for $Z^0 \rightarrow
\overline{q}q$\cite{PDG}.  It  therefore seems unlikely that particles can be
lighter than about 100 GeV and  still yield effects that are not extremely
suppressed\footnote{Herczeg,Kambor, Simonius and Wyler are currently
analyzing a model with a light boson (P.  Herczeg, private
communication).}.   One can argue that $\Lambda$ should be much larger, but
the conservative estimate we use is $\Lambda \sim$ 100 GeV.
 
To estimate the size of TVPC physical effects we must identify the 
lowest-dimensional TVPC operators in Eq. (\ref{e:L}).   There are no
gauge-invariant TVPC flavor-diagonal operators of dimension six or less
in standard-model  fields.\footnote{One that might appear to be
relevant is $\bar{\psi} \gamma_{\mu} i D_{\nu} G^{\mu\nu}_a \lambda_a \psi ~+
h.c.$, where $G^{\mu\nu}_a$ is the gluon field-strength tensor and the 
$\lambda_a$'s  are $SU(3)$ color matrices.  However, the equations of motion
$ iD^{\mu} \gamma_{\mu} \psi = m \psi$ can be used to show
that the operator vanishes up to
a surface term.  This statement, which  remains true with the addition of a
$\gamma_5$, implies that not all of the 81 dimension-six standard-model
operators in Ref.\cite{BW} are independent; the 10 involving fermions and
vectors can be reduced up to a surface term through equations of motion (with
Higgs scalars instead of explicit masses) to one of the operators involving
$\bar{\psi} \sigma_{\mu\nu} \psi'$, a field strength tensor, and a Higgs
scalar.} Dimension-six $SU(2)_L$-{\em non}invariant operators can  in 
principle
contribute to physical effects\cite{BL}, but these operators all contain weak 
gauge bosons that suppress their contributions to low-energy processes beyond
those of higher-dimensional  operators.
Therefore, it would appear that the largest local TVPC flavor-diagonal 
$SU(3)_C \times U(1)_Q$-invariant 
operators in the standard model have dimension seven.  All such
four-quark operators can  be written in the form\cite{Khrip}
\begin{eqnarray}  
\label{e:7} C_7 \left({1 \over \Lambda}\right)^3 \bar{q}_1
\gamma_5 D^{\mu} q_2 ~  \bar{q}_3 \gamma_5 \gamma_{\mu} q_4 ~ + h.c. \, ,
\end{eqnarray}  
where $C_7$ is a dimensionless constant expected to be of
order one, and  $q_1 = q_2,~ q_3 = q_4 \ne q_1$ or $q_1=q_4, ~q_2 = q_3 \ne
q_1$. [The Gordon decomposition can be used to replace $D^{\mu}$ by the
expression $\sigma^{\mu\nu} q_{\nu}$.]  There is also a quark-gluon-photon
operator of the form 
\begin{equation} \label{e:7p} C_7'\left({1 \over \Lambda}\right)^3 \bar{q}
\sigma_{\mu\nu} \lambda_a q~G^{\mu\rho}_a F^{\nu}_{\rho}~, 
\end{equation}
where $G^{\mu\rho}_a$ is the gluon field strength tensor and
$F^{\nu}_{\rho}$ is the electromagnetic field strength tensor.\footnote{
This operator was pointed out to us by D. Kaplan.}
The existence of these operators suggests that the largest TVPC
flavor-conserving term in the expansion of ${\cal L}_{eff}$ is in ${\cal
L}_3$, and that experimental effects should occur at a scale of order
$(p/\Lambda)^3$.  With conservative estimates for $p$ and $\Lambda$, this
naive result implies that the effects of any flavor-conserving TVPC operator
in low energy experiments must be suppressed by at least 
$10^{-6}$ relative to
strong interactions.   For our order
of magnitude estimates, we take the hadron-level interactions
to have roughly the same strength as the corresponding
quark-level interactions. This means, for instance, that $\bar{g_{\rho}}$, 
the
ratio of the TVPC $\rho N N$ coupling to the strong $\rho N N$ coupling, is
at most about  $10^{-6}$.  

Ref.\ \cite{CK} shows, however, that
the stringent experimental limits on the neutron electric dipole
moment (edm) imply that the effects of the dimension-seven 
operators are even
smaller. The diagram in Fig.\ 1 shows a potential contribution of
TVPC physics to the low-energy dimension-five quark-edm operator,
where 
the Z-boson exchange makes the diagram P violating. Matching
this diagram to the
effective theory valid at edm scales results in 
an estimate for the
coefficient $C_5$ in the dimension-five edm operator
\begin{equation}
\label{e:edm}
{C_5 \over \Lambda} \bar{q} \sigma_{\mu\nu} \gamma_5 q ~F^{\mu\nu}~
\end{equation}
on the order of
\begin{equation}
C_5 \sim {4 \pi \alpha \over (16 \pi^2)^2} C_7 \sim 4 \times 10^{-6} C_7\ \ .
\end{equation}
The measured limit\cite{EDM1,EDM2} on the neutron edm, $d_n/e \
\raisebox{-.25ex}{$\stackrel{<}{\scriptstyle \sim}$}\ 10^{-25}$ cm, gives
$C_5 \ \raisebox{-.25ex}{$\stackrel{<}{\scriptstyle \sim}$}\  5  \times
10^{-10}$ and therefore implies that $C_7 \ 
\raisebox{-.25ex}{$\stackrel{<}{\scriptstyle \sim}$}\  10^{-4}$. This 
indirect 
bound on the magnitude of $C_7$ reduces the
expected effects of the dimension-seven TVPC four-quark operators in flavor 
diagonal nuclear 
experiments to at most $10^{-10}$ the size of
strong effects, well beyond the reach of current or anticipated 
experiments.  The coefficent $C_7'$ in Eq.\ (\ref{e:7p}) can be bounded at a
similar level because the associated operator contributes to the  neutron edm
via another two-loop diagram containing a Z boson.

The suppression of the coefficients in the 
dimension-seven operators, however, does not translate into an equivalent 
suppression
of TVPC effects; larger ones can
come from operators of
dimension {\em eight}, provided they do not contribute significantly 
to the neutron edm.  An example of such an operator is 
\begin{equation}
\label{e:8}
{C_8 \over \Lambda^4}\bar{q} \gamma_{\mu} \gamma_5 q ~\bar{q} \gamma_{\nu} 
\gamma_5 \lambda_a q ~G^{\mu\nu}_a~,
\end{equation}
which represents the TVPC interaction of 
four quarks with a gluon.  The interactions giving rise to this operator 
could contribute to the neutron edm, 
but at a level much lower than in the case of the dimension-seven operators.  
The 
reason is that the dimension-eight operator does not itself flip chirality, 
and so when
inserted into a diagram like that of Fig.\ 1
(with the end of the gluon line attached to a  
quark line),
it must be accompanied by a quark mass on an external quark line to 
contribute
to the chirality-changing dimension-five operator in Eq.\ (\ref{e:edm}).  The 
coefficient $C_8$ can therefore assume its ``natural'' value of order one 
without generating a dipole moment larger than the measured limit (it cannot
be significantly enhanced, however, without doing so).  
The
suppression of operators that change chirality relative to those that 
do not is plausible; it may be that all chirality-changing
operators at low energies originate from fermion-mass insertions in the full
theory.  Such insertions would add factors such as $m_q/\Lambda$
(about $10^{-4}$ here) to the dimension-seven chirality-changing operators
without affecting the dimension-eight operator in Eq.\ (\ref{e:8}).
The reasonableness of this scenario, and in particular the lack of an 
experimental
constraint on $C_8$, means that the most natural bound on TVPC
effects in low-energy flavor-conserving experiments compared
to strong effects is not $10^{-6}$ or $10^{-10}$, but
$(m_p/\Lambda)^4 = 10^{-8}$.

We conclude as follows. 
Without knowing the source of physics beyond the standard model that
may induce TVPC couplings, we may use an effective Lagrangian
valid at low energies and estimate the size of its largest terms through
dimensional analysis.  Without any experimental input, we can
conclude that since $\bar{g_{\rho}}$ is generated only by operators of at 
least
dimension seven, it is very likely less than $10^{-6}$.  Existing 
limits on the neutron edm appear to constrain the effects of these operators,
however, so that the largest effects consistent with 
experiment arise from
operators of dimension eight. This results in a natural upper bound 
on $\bar{g_{\rho}}$ 
of
about $10^{-8}$.  
(If a measurement of a TVPC effect were obtained between 
$10^{-6}$ and $10^{-8}$, this would suggest that either
the high energy theory possesses some unknown symmetry, or
that accidental cancellations prevent the TVPV physics from
contributing to the neutron edm.)
Our estimates are conservative, and we conclude that a dramatic 
improvement in sensitivity is required for low-energy 
experiments to have a good chance of seeing TVPC effects.

This work was supported in part by the U.S.  Department of Energy under
Grants DE-FG05-94ER40827; DE-FG05-85ER-40219, Task B; and
DE-FG05-90ER40592.  We thank the authors of Ref.\ \onlinecite{Herczeg} for
a pre-publication copy of their work and Xiangdong Ji, David
Kaplan, Yulik Khriplovich, David London,
Berndt M\"uller, Michael Musolf, Peter van
Nieuwenhuizen, Martin Savage, and Mark Wise for informative discussions.

\begin{figure}
\epsfxsize=5cm
\hfil\epsfbox{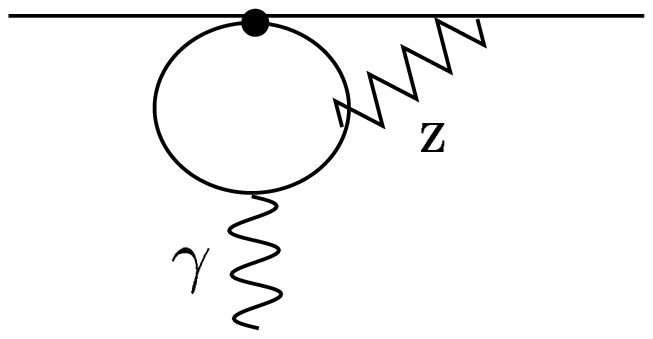}\hfill
\caption{One diagram that could give rise to a neutron edm at
matching.  Solid lines are quarks, the jagged line is a Z-boson, and
the wavy line is a photon.  The solid dot indicates the insertion
of TVPC physics.}
\label{diag}
\end{figure}


\begin{references}

\bibitem{KM}
M.~Kobayashi and K.~Maskawa, Prog.~Theor.~Phys.~{\bf 49}, 652 (1973).

\bibitem{gould} 
See {\sl Time Reversal Invariance and Parity Violation in Neutron
Reactions}, ed. C.R.~Gould, D.~Bowman, and Y.P.~Popov (World Scientific,
Singapore, 1994), and references therein.

\bibitem{gamma}
N.K.~Cheung, H.E.~Henrikson, and F.~Boehm, Phys.~Rev.~{\bf C24}, 
620 (1981). 

\bibitem{Koster} 
J.E.~Koster et al., Phys.~Lett.~{\bf B267}, 23 (1991).

\bibitem{db}
E.~Blanke et al., Phys.~Rev.~Lett.~{\bf 51}, 355 (1983).

\bibitem{beta}
A.L.~Hallin et al., Phys.~Rev.~Lett.~{\bf 52}, 337 (1984).

\bibitem{Simonius}
M.~Simonius, Phys.~Lett.~{\bf B58}, 147 (1975).


\bibitem{kind1}
J.~Engel, C.R.~Gould, and V.~Hnizdo, Phys.~Rev.~Lett.~{\bf 73}, 3508 
(1994).

\bibitem{kind2}
W.C.~Haxton, A.~H\"oring, and M.~Musolf, Phys.~Rev.~{\bf D50}, 3422 
(1994).

\bibitem{Herczeg}
P.~Herczeg, J.~Kambor, M.~Simonius and D.~Wyler, Parity-Conserving 
Time-Reversal 
Violation in Flavor-Conserving Quark-Quark Interactions, to be published.
 
\bibitem{CK} 
R.S.~Conti and I.B.~Khriplovich, Phys.~Rev.~Lett.~{\bf 68}, 3262 
(1992).

\bibitem{Leff}
K.~Symanzik, Comm.~Math.~Phys. {\bf 34}, 7 (1973),\\
S.~Weinberg, Phys.~Lett.~{\bf 91B}, 51 (1980),\\
H. ~Georgi, {\it Weak Interactions and Modern Particle Theory.}
Benjamin/Cummings, Menlo Park, CA. (1984)\\
For a review, see H. Georgi, Ann. Rev. Nucl. Prt. Sci. {\bf 43}, 209 
(1993).

\bibitem{har} A. Manohar and H. Georgi, Nucl. Phys. {\bf B234}, 189 (1984);
H. Georgi and L. Randall, Nucl. Phys. {\bf B276}, 241 (1986);
H. Georgi, Phys. Lett. {\bf298B}, 187 (1993).

\bibitem{PDG}
Particle Data Group, Review of Particle Properties, Phys. Rev. {\bf D50}. 
Part I, 1173 (1994)

\bibitem{BW}
W.~Buchm\"uller and D.~Wyler, Nucl.~Phys.~{\bf B268}, 621 (1987),\\
W.~Buchm\"uller, B.~Lampe, and N.~Vlachos, Phys.~Lett. {\bf B197}, 
379 (1987).  

\bibitem{BL}
C. P. ~Burgess and D. ~London, Phys. Rev. {\bf D48}, 4337 (1993)

\bibitem{Khrip}
I.B.~Khriplovich, Nucl.~Phys.~{\bf B352}, 385 (1991).

\bibitem{EDM1} K.F. Smith et al., Phys. Lett. {\bf 234B}, 191 (1990).

\bibitem{EDM2} I.S. Altarev et al., Phys. Lett. {\bf 276B}, 242 (1992).

\end{references}
\end{document}